# Deep Learning for Dermatology: An Innovative Framework for Approaching Precise Skin Cancer Detection

Mohammad Tahmid Noor
*Department of Computer Science and Engineering*
*East West University*
Dhaka, Bangladesh
tahmidnoor770@gmail.com

B. M. Shahria Alam
*Department of Computer Science and Engineering*
*East West University*
Dhaka, Bangladesh
bmshahria@gmail.com

Tasmiah Rahman Orpa
*Department of Computer Science and Engineering*
*East West University*
Dhaka, Bangladesh
tasmiahorpa376@gmail.com

Shaila Afroz Anika
*Department of Computer Science and Engineering*
*East West University*
Dhaka, Bangladesh
anikaafroz2002@gmail.com

Mahjabin Tasnim Samiha
*Department of Computer Science and Engineering*
*East West University*
Dhaka, Bangladesh
tasnimsamiha864@gmail.com

Fahad Ahammed
*Department of Computer Science and Engineering*
*East West University*
Dhaka, Bangladesh
fahadahbd@gmail.com

*Abstract*— Skin cancer can be life-threatening if not diagnosed early, a prevalent yet preventable disease. Globally, skin cancer is perceived among the finest prevailing cancers and millions of people are diagnosed each year. For the allotment of benign and malignant skin spots, an area of critical importance in dermatological diagnostics, the application of two prominent deep learning models, VGG16 and DenseNet201 are investigated by this paper. We evaluate these CNN architectures for their efficacy in differentiating benign from malignant skin lesions leveraging enhancements in deep learning enforced to skin cancer spotting. Our objective is to assess model accuracy and computational efficiency, offering insights into how these models could assist in early detection, diagnosis, and streamlined workflows in dermatology. We used two deep learning methods DenseNet201 and VGG16 model on a binary class dataset containing 3297 images. The best result with an accuracy of 93.79% achieved by DenseNet201. All images were resized to 224×224 by rescaling. Although both models provide excellent accuracy, there is still some room for improvement. In future using new datasets, we tend to improve our work by achieving great accuracy.

I. INTRODUCTION

Skin cancer manifests in different types, known as the part of the most concerning and common forms of cancer on a global basis. Among the differences in skin cancer, melanoma is the most aggressive due to its potential to spread rapidly. Prolonged exposure to ultraviolet radiation is mainly responsible for the condition arising. This condition leads to abnormal cell growth in the skin. To substantially improve survival rates, early detection is critical. The challenging part is to identify benign or malignant lesions, due to the visual differences and overlapping characteristics of these lesions.

Skin cancer arises from abnormal skin cell changes that proliferate uncontrollably. These skin cancers form tumours that can be either malignant or benign. Basal cell carcinoma and squamous cell carcinoma are two other varieties of skin cancer that tend to grow slowly and rarely spread. However, skin cancer is a leading form of cancer worldwide that affects the areas vulnerable to ultraviolet radiation, such as the face, arms, legs, etc.

UV rays are mainly responsible for spreading skin cancer from the sun or synthetic sources like tanning beds. Damaging the DNA in skin cells leads to mutations and causes cells to grow immensely. Excessive sun exposure without protection, a sluggish immune system, a family history of skin cancer, etc can also increase the possibility of one's skin cancer.

Minimizing disclosure to UV radiation is the main prevention of skin cancer [1]. Wearing vigilant clothing, using broad-spectrum sunscreen with immense SPF, seeking shade during excessive sunlight and heat, etc cloud be helpful in terms of protecting every individual. Regular dermatologist screenings are crucial for catching early signs of any kind of dermatological concerns. To significantly dwindle the risk of abnormal skin growth, every individual should adopt these preventions as per public health campaign recommendations.

In the realm of medical imaging, especially classifying skin lesions, Convolutional neural networks (CNN), particularly VGG16 and DenseNet201 demonstrated profound efficiency [2]. To extract and analyze patterns, the models are designed, and they can analyze intricate patterns from high-resolution dermoscopic images. For their architectural depth and capability in feature extraction, VGG16 and DenseNet201 are profoundly renowned. This quality allows them to process the complex texture, color, and characteristics that distinguish healthy skin.

In the era of image classification, VGG16, a 16-layer deep CNN set a high standard due to its simplicity and high performance [3]. Sequential convolutional layers with a uniform kernel size are there in the architecture of this model that allows it to maintain computational efficiency and capture essential visual features. VGG16 is mainly stronger in hierarchical feature extraction, which enables learning low-level edges and textures in early layers. VGG16's structure is built in a way that enables one to discern the subtle textual differences between benign and malignant lesions by focusing on details within the image.

Another model that represents a more advanced CNN architecture known as DenseNet201, emphasizes extensive interconnections between layers. In traditional models, each layer passes information to the next while DenseNet201 connects each layer to every other layer. It promotes feature reuse and reduces the number of parameters that are required. DenseNet201 seems to capture complex, nuanced patterns within dermoscopic images, enhancing its ability to extricate itself amidst different skin irregularities. This model can analyze intricate structures within malignant spots, with varied pigmentation and asymmetric shapes. Another paper combined U-Net with an improved version of MobileNet-V3 architecture to make it more precise to diagnosing cancer which was enhanced through hyperparameter optimization [4].

The development and application of these models can emphasize the transmute potential of AI in healthcare. These models promise us an early detection and meticulous diagnosis of skin cancer, which can become routine and improve patient outcomes globally.

In skin cancer diagnosis, both VGG16 and DenseNet201 models have automated formidable tools, offering distinct advantages to cater to different clinical needs. VGG16 architecture is simple, and it has moderate resource demand which makes it ideal for fast, scalable applications. The architecture's speed and efficiency are paramount. On the other hand, DenseNet201 provides us with deeper feature insights which is particularly effective in time requiring high sensitivity. Based on clinical requirements, combining these models, diagnostic accuracy can be optimized by the healthcare providers. The development and application of these models reinforce the transformative possibility of AI in healthcare which promises us a future where early detection and meticulous diagnosis of skin cancer can be routine and provide outcomes on a global scale. Leveraging a larger and more diverse dataset, our approach regulates the potential to attain superior diagnostic accuracy, thereby amplifying its robustness, generalizability, and reliability in being prudent to intricate and subtle patterns across a broad spectrum of skin cancer cases. This advancement underscores the stowage of our method to transfigure diagnostic practices, paving the way for more rigorous, versatile, and stirring implications in the detection and management of skin cancer.

II. RELATED WORKS

For better understanding, we have gathered some related works that can help us to understand more clearly the comparison of our proposed work.

Hasan et al. make an ingenious approach to skin cancer diagnosis by leveraging convolutional neural networks to automate the detection process [5]. Their study introduces a CNN architecture that processes dermoscopic images to classify skin lesions as benign or malignant. The model achieves a notable accuracy of 89.5% by segmenting images and extracting defining features. The efficiency and precision of CNNs were emphasized in this paper in medical imaging. An alternative pathway was proposed to expedite diagnosis and reduce diagnostic errors in skin cancer detection.

Battle et al. handled Siamese Neural Networks (SNNs) to classify surface lesions and ascertain the novel league of skin cancer adopting clinical and dermoscopic pictures [6]. Their leading-edge approach achieved a classification accuracy of 74.33% on clinical images and 85.61% on dermoscopic images. Their accuracy level evinces the model's adeptness at extricating between highly similar lesion classes. To address the intricacy inherent in dermatological diagnosis, peculiarly when novel or rare lesion categories emerge, by showcasing the potentials of Siamese Neural Networks (SNNs) in this study. This model provides a robust framework by learning discriminative features between image pairs and successfully identifies skin cancer types with greater precision.

Krohling et al. utilize a Convolutional based deep learning model to train pictures, which are mingled with lesion specific data and patient demographics [7]. Their study compassed a balanced accuracy of 85%, which integrated into a smartphone-based application, and they steered at making skin cancer screening more attainable and convenient. Their integration of deep learning into a mobile platform that highlights its potential in enhancing the reach and persuasiveness of dermatological assessments. Image data and patient context were leveraged, and the CNN based approach delivers reliable and susceptible classification. This application demonstrates the power of combining deep learning and mobile technology to democratize healthcare, fostering early detection.

Shaikh et al. emphasize the role of CNNs transformation diagnostic practices and explore the potential of AI-driven solutions for skin cancer detection [8]. Their study presents a CNN based approach that automates lesion analysis and reduces the reliance on manual examination. By isolating key features and segmenting dermoscopic images, the model effectively differentiates amongst benign and malignant skin lesions. Using the CNN architecture, they got an impressive accuracy of 91% which demonstrates CNN's capacity to provide reliable and timely diagnoses.

Ganthya's work demonstrates a novel CNN model that is designed to advance skin cancer which enhances image classification technique [9]. Their model achieves an excellent accuracy of 91% by leveraging advanced data preprocessing methods which also include data augmentation. They have emphasized reducing diagnostic errors and present their model as a promising tool to assist doctors in the prompt recognition and verdict of skin conditions.

III. METHODOLOGY

In the methodology section, we provided a detailed description of our proposed work. We applied a deep learning model to our preprocessed dataset. For VGG16, we used a learning rate of 0.0002 and optimized using the Adam optimizer for efficient convergence. The batch size was set to 20 by balancing memory and stable gradient updates. By implementing early stopping to prevent overfitting, we trained the model for 108 epochs. For DenseNet201, Epoch 80, Patience 15, Learning rate 0.0001, and batch size 20 have been used. Some dataset enrichment methods such as rotation, flipping, and contrast alteration helped improve generalization. We also fine-tuned the hyperparameters and introduced dropout layers to enhance model robustness. To extract deep features using transfer learning, we trained the deep CNN model. We use this feature to recognize the patterns. To shape, color, print, and appearance to represent an object, we use the deep CNN algorithm. There are different layers in the deep CNN model to calculate the dot product of weights and pass the input image in the first layer. In the ReLu layer, inactive neurons were done by pooling and removing. Softmax is being used to classify the features that are computed. We also used VGG16 which is a pre-trained deep CNN model to extract its features. Fig. 1 shows the proposed methodology.

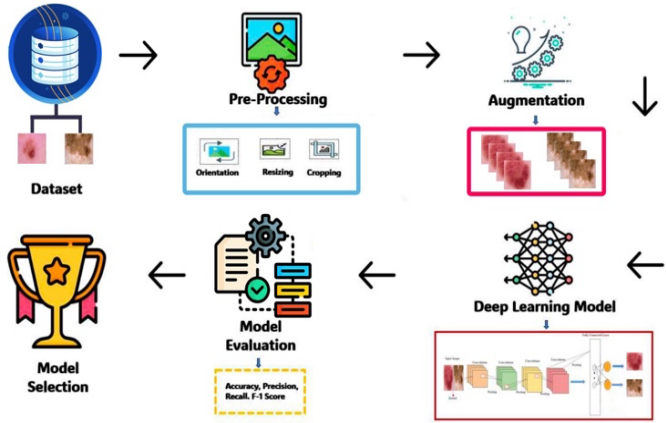

Fig. 1. Methodology

### A. Dataset

We have used the proposed models for disease detection. For this research, we use two types of disease datasets that are benign (healthy) and malignant (infected) skin spots that have 1800 and 1497 images [10]. We collected the dataset from online sources. We found some imbalanced distribution, with malignant cases being fewer than benign. We have done some preprocessing, including reseizing, normalizations, data augmentation, and noise reduction, to enhance model performance. Faced some challenges including dataset imbalance and varying image quality which we addressed using oversampling, augmentation, and filtering of low-quality images. The proposed methods were simulated on a MacBook Air M2 15-inch. We have shown a sample input of the dataset in the following Fig. 2. This balanced and comprehensive dataset forms the substratum of our analysis, enabling the model to congruously learn subtle distinctions between healthy and cancerous skin conditions, thereby driving agile and flawless predictions.

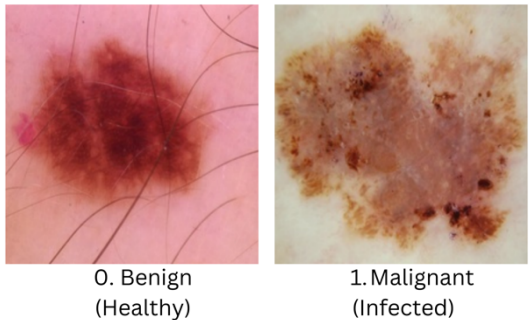

Fig. 2. Sample inputs of the skin cancer

In the image 0, skin cancer named benign is shown. Benign skin spots are generally harmless and noncancerous. It naturally is consistent in color and shape, slightly raised texture. On the other hand, Malignant skin spots are shown in the pictures 1. It is potentially cancerous. This cancer can be seen as asymmetry and irregular borders, and it may also itch or bleed. These discernible patterns amplify the dataset's ability to convey critical diagnostic indications, providing a panoramic keystone for training a model.

### B. VGG-16

In the era of image classification, VGG16 is so strong that has done an impressive job. VGG16, a convolutional neural network, is pre-trained model based on a massive dataset. After extracting the features using a pre-trained architecture, fine-tune the entire model. Expanding the depth of the network, VGG16 joins more convolutional layers while using very small convolutional filters in all layers. Hyperparameter tuning for VGG16 is shown in Table 1. In Fig. 3 the structure of the VGG16 engineering is shown.

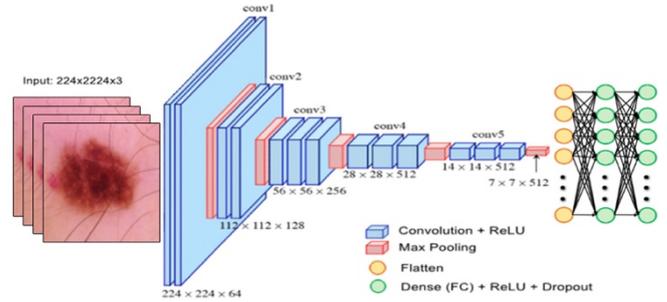

Fig. 3. Structure of the model VGG16

TABLE I HYPERPARAMETER TUNING FOR VGG16

| Batch size | 20 | Loss function | binary_crossentropy |
|---|---|---|---|
| Learning rate | 0.0002 | Number of epochs | 108 |
| Optimizer | Adam | Patience | 20 |

### C. DenseNet201

DenseNet201, a convolutional neural network, establishes connections between layers. This model has such layers as convolutional networks, pooling, and fully connected layers. Utilizing the bottleneck layers and transition layers improves computational efficiency. This model is a powerful tool for detecting skin cancer. Hyperparameter tuning for DenseNet201 is shown in Table 2.

TABLE II HYPERPARAMETER TUNING FOR DENSENET201

| Batch size | 20 | Loss function | binary_crossentropy |
|---|---|---|---|
| Learning rate | 0.0001 | Number of epochs | 62 |
| Optimizer | Adam | Patience | 15 |

An advanced deep learning model like DenseNet201 promises to enhance crop health and productivity in agricultural sectors. Fig. 4 shows the architecture of DenseNet201.

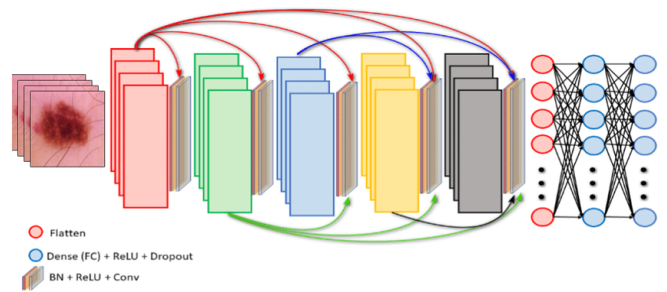

Fig. 4. Structure of the model DenseNet201

## IV. EXPERIMENTAL RESULT AND ANALYSIS

### A. VGG-16

In this process, 87.49% accuracy has been successfully gained by the VGG16 model. It also succeeded for 108 epochs. 330 seconds per step was taken during the training phase and we applied 108 training epochs. In Fig. 5, 108 epochs are displayed. Accuracy for VGG16 as classification wise has shown in Table 3.

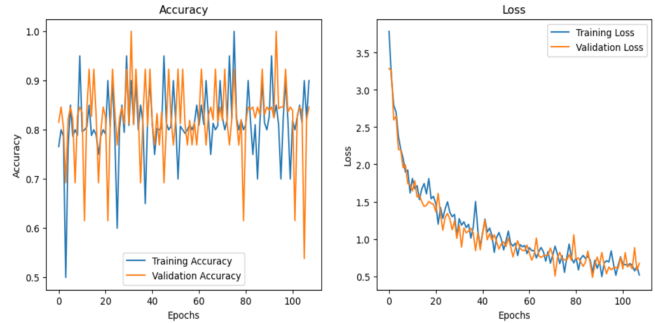

Fig. 5. VGG16 model – quality metrics evaluation during training: (a) accuracy (b) loss

TABLE III CLASSIFICATION WISE ACCURACY TABLE FOR VGG16

| Classes | Precision | Recall | F1-score | Support | Accuracy |
|---|---|---|---|---|---|
| Benign | 0.95 | 0.78 | 0.86 | 180 | 0.86 |
| Malignant | 0.78 | 0.95 | 0.86 | 149 | |

Fig. 6 shows the performance of the VGG16 algorithm which has been highlighted in this confusion matrix in classifying benign and malignant skin spots. With 140 benign and 142 malignant cases correctly classified by this model which indicates high accuracy. However, 40 benign cases were classified as malignant, and 7 malignant cases were misclassified as benign which can lead to unnecessary tests. Overall the model is effective and could benefit from reducing false positives and improving specificity.

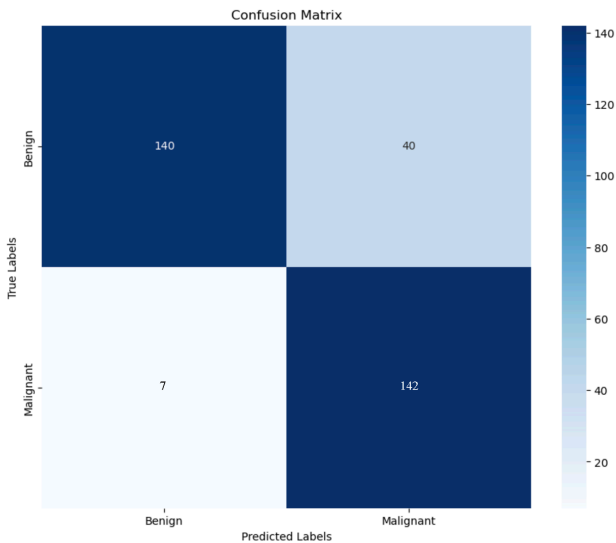

Fig. 6. Confusion matrix of VGG16 model

### B. DenseNet201

The DenseNet201 model has successfully gained a training accuracy of 93.79% after training for 80 epochs. 374 seconds per step has taken by the training phase and we applied 80 training epochs. The epochs are displayed in Fig. 7 and the classification-wise accuracy is shown in Table 4.

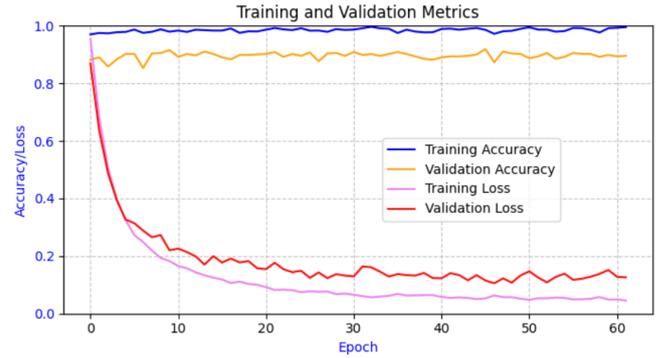

Fig. 7. DenseNet201 model – quality metrics evaluation during training: (a) accuracy (b) loss

TABLE IV CLASSIFICATION WISE ACCURACY TABLE FOR DENSENET201

| Classes | Precision | Recall | F1-score | Support | Accuracy |
|---|---|---|---|---|---|
| Benign | 0.98 | 0.89 | 0.94 | 180 | 0.93 |
| Malignant | 0.88 | 0.98 | 0.93 | 149 | |

In Fig. 8, This confusion matrix demonstrates improved classification by the model, with 161 benign and 146 malignant cases correctly identified. In this model, benign is misclassified as malignant which is only 19 and the false negative count which is malignant as benign is very low only 3. The overall enhancement is reflected by this model in both sensitivity and specificity. This shows that the model is more accurate in identifying both benign and malignant cases.

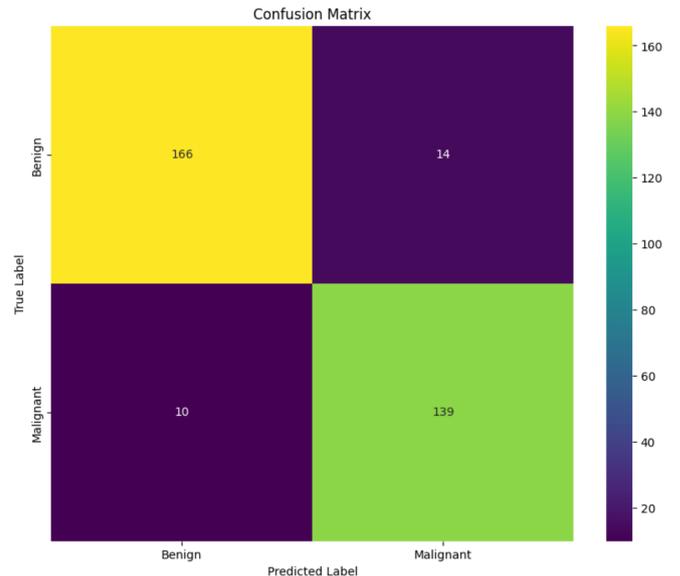

Fig. 8. Confusion matrix of DenseNet201 architecture

### C. SHAP

Fig. 9 shows the skin cancer identification by using DenseNet201. SHAP mainly assigns value to the feature and explains how much each contributes to the prediction. A prominent feature here is Grad-CAM. The following figure shows a Grad-CAM output where the heatmap highlights the most influential regions that contributed to the deep learning model's prediction. The picture visualizes that the heatmap highlights the spot on skin which belongs from the benign class and confirms that the model identifies these areas as important features. In the heatmap the warm red color areas indicate areas of greater significance. By this, the model makes decisions based on relevant visual cues.

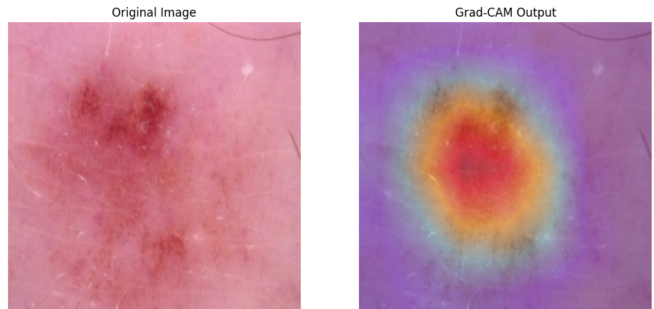

Fig. 9. Skin Cancer Identification using SHAP

## V. COMPARISON

The training time between the two models VGG16 was taken 9 hours 23 minutes and the DenseNet201 was taken 8 hours 31 minutes. The training and test accuracy are compared in Table 5. The two models have gained an excellent accuracy of 87.49% and 93.79%. The training and the test accuracy of each model have different numbers of training epochs. Despite having 93.79% accuracy, it struggles with ambiguous lesions where the two classes feature overlap and misclassification happens. Generalizations to real-world cases are affected because of the dataset's limited diversity. We aim to use more advanced augmentation techniques for better diagnostic accuracy.

TABLE V ACCURACY COMPARISON BETWEEN THE TWO MODELS

| Algorithm | VGG16 | DenseNet201 |
| --- | --- | --- |
| Training accuracy | 87.49 | 93.79 |
| Test accuracy | 86.93 | 93.31 |

To showcase our work and to prove that our method was more effective than the others we gathered some comparisons with some previous works shown in Table 6. In terms of accuracy, various skin cancer classification methods are compared. Using convolutional neural networks (CNNs), Hasan et al. achieved 89.50% accuracy [5]. Wu attained 83.10% accuracy with a ResNet-50 model [11], and Houssein's deep CNN approach achieved 87.90% [12], also under 90%. In contrast, Battle et al. SNNs to achieve 85.61% [6] demonstrating improved performance. Krohling et al. an achieved accuracy of 85% by using CNN architecture [7]. On the other hand, using VGG16 architecture, we can achieve 87.49% accuracy and DenseNet201 gives us an accuracy of 93.79% which shows our model works better in terms of better accuracy and detection of skin cancer. The F1 score ensures balanced performance across two classes. The AUC-ROC measured its ability to make a difference between benign and malignant cases. By analyzing the precision and recall, we minimize false negatives in malignant detection. Our approach has yielded unparalleled outcomes, surpassing all other methods, thereby underscoring its exceptional efficacy and distinguished superiority in delivering remarkable results.

TABLE VI COMPARISON OF THE PROPOSED MODEL WITH PREVIOUS WORKS

| Author | Method | Accuracy (%) |
| --- | --- | --- |
| Hasan et al. [5] | Convolutional neural networks | 89.50 |
| Battle et al. [6] | Siamese Neural Networks | 85.61 |
| Wu [11] | ResNet-50 | 83.10 |
| H. Houssein [12] | Deep Convolutional Neural Network | 87.90 |
| Krohling et al. [7] | CNN | 85.00 |
| Proposed model | CNN VGG16 | 87.49 |
| | DenseNet201 | 93.79 |

## VI. CONCLUSION

Skin cancer ranks among the most ubiquitous and expeditiously growing cancers worldwide, posing a significant health challenge, especially due to its increasing rate from influences like persistent UV exposure. Early and accurate diagnosis is crucial, particularly for malignant varieties like melanoma, where timely intervention can greatly expand the responses from patients. In this article, we explored the efficiency of deep neural learning algorithms, specifically VGG16 and DenseNet201 to classify skin abnormalities as either malignant or benign. The articles were thoroughly qualified and assessed on a dataset to evaluate their efficacy in accurately identifying skin cancer, a fundamental application within dermatology and oncology. VGG16 achieved an accuracy of 87.49%, while DenseNet201 displayed outstanding performance with an accuracy of 93.79%, highlighting its potential for enhanced diagnostic accuracy in clinical approaches. The method's success in attaining high accuracy after 80 epochs solidifies its role as a groundbreaking advancement in the field, offering a scalable and viable solution for prompt detection. Harnessing the synergy of DenseNet201's dense connectivity and VGG16's proven efficiency, this method demonstrates a new benchmark in the quest for navigable and comprehensive cancer diagnostics. We hope to use vision transformers in the future for improved feature extraction, integrating explainable AI for better interpretability. Thus, the proposed approach is not merely a technical achievement but a step forward in democratizing access to accurate skin cancer diagnostics globally. Finally, our study sheds light on the critical role of convolutional neural networks in dermatological diagnostic assessments. These insights reinforce the transformative possibility of deep learning in medical diagnostics, offering a pathway toward more precise, accessible, and life-saving tools for skin cancer detection.